\RequirePackage{lineno}
\documentclass[aps,prd,twocolumn,showpacs,amsmath,amssymb]{revtex4-1}
\usepackage{overpic}
\usepackage{graphicx}
\usepackage{epsfig}
\usepackage{dcolumn}
\usepackage{bm}
\usepackage{rotating}
\usepackage{multirow}
\usepackage{booktabs}
\usepackage{appendix}
\usepackage{subfigure}
\usepackage{hhline}
\usepackage{amssymb}
\usepackage{lineno}
\usepackage{hyperref}
\usepackage{mathrsfs}
\usepackage{verbatim}
\usepackage{amsmath}
\usepackage{amssymb}
\usepackage{amsfonts}
\usepackage{amsbsy}
\usepackage{float}
\usepackage{natbib}

\hypersetup{colorlinks,linkcolor=blue,anchorcolor=blue,citecolor=blue}
\include{def-com}

\def \ee {e^+e^-}

\def \half {{1\over 2}}
\def \ee {e^+e^-}

\def \cP {\mathcal{P}}
\def \cW {\mathcal{W}}

\newcommand{\eq}[1]{Eq.~\eqref{eq:#1}}

\hypersetup{colorlinks,linkcolor=blue,anchorcolor=blue,citecolor=blue}
\include{def-com}

\let\oldequation\equation
\let\oldendequation\endequation

\renewenvironment{equation}
  {\linenomathNonumbers\oldequation}
  {\oldendequation\endlinenomath}

\def\ee{e^{+}e^{-}}

\def \miss2{M_{\rm miss}^{2}}

\def \romanOne   {\uppercase\expandafter{\romannumeral1}}
\def \romanTwo   {\uppercase\expandafter{\romannumeral2}}
\def \romanThree {\uppercase\expandafter{\romannumeral3}}
\def \romanFour  {\uppercase\expandafter{\romannumeral4}}
\def \romanFive  {\uppercase\expandafter{\romannumeral5}}
\def \romanSix   {\uppercase\expandafter{\romannumeral6}}
\def \romanSeven {\uppercase\expandafter{\romannumeral7}}
\def \romanEight {\uppercase\expandafter{\romannumeral8}}
\def \romanNine {\uppercase\expandafter{\romannumeral9}}

\newcommand{\lambdacp}{\Lambda_{c}^{+}}

\newcommand{\sigmode}[1]{
	\ifnum#1=1
	\lambdacp \rightarrow n K_{S}^{0} \pi^{+}
	\else
	\ifnum#1=2
	\lambdacp \rightarrow n K_{S}^{0} K^{+}
	\fi
	\fi
}

\newcommand{\alphaXi}{\alpha_{\Xi^+_c}}
\newcommand{\betaXi}{\beta_{\Xi^+_c}}
\newcommand{\gammaXi}{\gamma_{\Xi^+_c}}

\begin{document}

\title{\boldmath Study of $P$ and $CP$ symmetries in $\Xi^+_c\rightarrow  \Xi^-\pi^+\pi^+$ at electron-positron collider}

\author{Yunlu Wang}
\email{yunluwang20@fudan.edu.cn}
\affiliation{Key Laboratory of Nuclear Physics and Ion-beam Application (MOE) and Institute of Modern Physics, Fudan University, Shanghai, China 200433}

\author{Yunlong Xiao}
\email{xiaoyunlong@fudan.edu.cn(Corresponding author)}
\affiliation{Key Laboratory of Nuclear Physics and Ion-beam Application (MOE) and Institute of Modern Physics, Fudan University, Shanghai, China 200433}

\author{Pengcheng Hong}
\email{hongpc@ihep.ac.cn}
\affiliation{College of Physics, Jilin University, Changchun, China 130012}
\affiliation{Institute of High Energy Physics, Beijing 100049}

\author{Ronggang Ping}
\email{pingrg@ihep.ac.cn}
\affiliation{Institute of High Energy Physics, Beijing 100049, People's Republic of China and University of Chinese Academy of Science, Beijing 100049, China}

\date{\today}

\begin{abstract}

Symmetry studies represent one of the most promising frontiers in particle physics research. This investigation focuses on exploring $P$ and $CP$ symmetries in the charm system through the measurement of asymmetry decay parameters in the three-body decay of $\Xi_c^{+}$. Incorporating electron and positron beam polarization effects and utilizing the helicity formalism, we characterize the decay of $\Xi_c^{+}$ and its secondary hyperons through asymmetry decay parameters. The complete angular distribution formula for these decays has been systematically derived. Our study evaluates the sensitivity of the asymmetry parameters for the $\Xi_c^{+} \to \Xi^{-}\pi^{+}\pi^{+}$ decay channel under various data sample sizes and beam polarization scenarios. These findings establish a robust theoretical framework for future experimental studies at the STCF, providing valuable insights for symmetry investigations in the charm sector.

\end{abstract}

\maketitle
\oddsidemargin -0.2cm
\evensidemargin -0.2cm

\section{\boldmath Introduction}

Charge parity ($CP$) violation is one of the three fundamental conditions required to explain the matter-antimatter asymmetry observed in the universe~\cite{Sakharov:1967dj}. In the Standard Model of particle physics, the quark dynamics are descried by the Cabibbo-Kobayashi-Maskawa (CKM) mechanism~\cite{Cabibbo:1963yz,Kobayashi:1973fv}. 
While the $CP$ violation observed in the decays of $K$, $B$, and $D$ mesons is well described by the CKM mechanism~\cite{Christenson:1964fg, Belle:2001qdd, BaBar:2001ags, Belle:2004nch, BaBar:2004gyj, LHCb:2019hro,Li:2019hho}, it remains insufficient to account for the observed matter-antimatter asymmetry in the universe~\cite{Peskin:2002mm}. Given that the observable universe is predominantly composed of baryons, studying $CP$ violation in baryon decays is crucial for understanding this asymmetry.
Recent experimental advancements have provided compelling evidence for $CP$ violation in $\Lambda_{b}$ baryons.
For instance, the evidence of $CP$ violation in  $\Lambda_{b} \to \Lambda K^+ K^-$~\cite{LHCb:2024yzj} decay and the first observation of $CP$ violation in the four-body decay of $\Lambda_{b} \to p K^-\pi^+\pi^-$~\cite{LHCb:2025ray} have been reported at the LHCb experiment.
However, current research on $CP$ violation primarily focuses on $\Lambda_{b}$ decays, making it challenging to directly compare theoretical predictions involving baryon decays without $b$-quarks. This highlights the need to explore $CP$ violation in other baryonic systems.
The two-body decay processes of baryons can be parameterized using decay parameters $\alpha$ and $\bar{\alpha}$. Parity violation in baryons and anti-baryons is indicated by non-zero values of $\alpha$ and $\bar{\alpha}$~\cite{Lee:1957he}. Experimentally, $CP$ symmetry is tested by comparing the decay parameters of baryons and anti-baryons through the $CP$-odd observable $\mathcal{A}_{CP} = (\alpha + \bar{\alpha}) / (\alpha - \bar{\alpha})$~\cite{BESIII:2021ypr,BESIII:2020fqg,BESIII:2023sgt,BESIII:2022qax,BESIII:2025fre}. Here, $\alpha$ and $\bar{\alpha}$ represent the decay parameters of the baryon and anti-baryon, respectively. A non-zero measurements of $\alpha$ and $\mathcal{A}_{CP}$ signify parity ($P$) and $CP$ violations~\cite{He:1999bv, Tandean:2003fr, Salone:2022lpt, He:2022bbs, Goudzovski:2022vbt}. While two-body decays of baryons have been extensively studied, identifying new sources of $CP$ violation in baryon decays without $b$-quarks is essential to unravel this mystery and represents a critical frontier in $CP$ violation research.

The Cabibbo-favored three-body decay of the charmed baryon, $\Xi_c^+ \to \Xi^- \pi^+\pi^+$, involves both external and internal $W$-emission modes, with decay amplitudes comprising factorizable and non-factorizable contributions. The total branching fractions of $\Xi_c^+ \to \Xi^- \pi^+\pi^+$ and its isospin channel $\Xi_c^+ \to \Xi^0 \pi^0\pi^+$ are approximately 10\%, making them the most probable decay modes~\cite{ParticleDataGroup:2020ssz}. Consequently, these channels are expected to yield abundant experimental data, facilitating the investigation of $P$ and $CP$ violations at future facilities such as the Super Tau-Charm Facility (STCF)~\cite{Achasov:2023gey}. 

The parameters for three-body decays cannot be directly described by the Lee-Yang formalism~\cite{Lee:1957qs}. Therefore, it is necessary to define analogous decay parameters through decay amplitudes to study $P$ and $CP$ violations. 
In electron-positron collisions, the threshold energy for the process $e^+e^- \to \Xi_c^+\bar{\Xi}_c^-$ is above 4.94 GeV. Both the upgraded BESIII experiment and the future STCF are expected to produce $\Xi_c^+$ and $\bar{\Xi}_c^-$ pairs. Precision measurements of the decay parameters for $\Xi_c^+ \to \Xi^- \pi^+\pi^+$ and its isospin process $\Xi_c^+ \to \Xi^0 \pi^0\pi^+$ will provide critical insights to validate or exclude existing theoretical models and test isospin symmetry through comparative decay parameters.

The amplitude for decay $\Xi^+_c(\lambda_1) \rightarrow \Xi^-(\lambda_2) \pi^+\pi^+$ would indicate all information about spin and helicity of
the particles. In helicity formalism, the amplitude is model-independent and expressed by $B^J_\mu(\lambda_2,0,0)$ which has the sum of resonances in \eq{B-int}. Here $J=\frac12$ is the spin of mother particle, $\mu$ is its eigenvalue in the body-fixed z-axis that can be $\frac12$ or $-\frac12$, and $\lambda_2=\pm\frac12$ is the helicity of $\Xi^-$. Then there are four degrees of freedom
$B^{\frac12}_{\frac12}(\frac12,0,0), B^{\frac12}_{-\frac12}(\frac12,0,0), B^{\frac12}_{\frac12}(-\frac12,0,0)
$, 
and
$
B^{\frac12}_{-\frac12}(-\frac12,0,0)$,
which are denoted by $B^{\mu}_{\lambda_2}$ for simplification.

Similarly, the production of $\Xi^+_c$ and decay of $\Xi^-$ are also written in helicity formalism, and the corresponding amplitudes are listed in Tab.~\ref{tab:decay}. If the parity is conserved~\cite{0025}, the four amplitudes should satisfy the relations $B_{\frac{1}{2}}^+ = -B_{-\frac{1}{2}}^+$ and $B_{\frac{1}{2}}^- = B_{-\frac{1}{2}}^-$. However, the weak decay process $\Xi^+_c \rightarrow \Xi^- \pi^+\pi^+$ allows for parity violation. To characterize parity violation in this process, three asymmetry parameters can be defined by following formula:
\begin{align}\label{eq:abc}
\alpha_{\Xi^+_c}
= &
\frac{|B^+_{\frac12}|^2 - |B^+_{\!-\frac12}|^2}{|B^+_{\frac12}|^2 + |B^+_{\!-\frac12}|^2}
\,, \nonumber \\
\beta_{\Xi^+_c}
= &
\frac{|B^-_{\frac12}|^2 - |B^-_{\!-\frac12}|^2}{|B^-_{\frac12}|^2 + |B^-_{\!-\frac12}|^2}
\,, \nonumber \\
\gamma_{\Xi^+_c}
= &
\frac{|B^+_{\frac12}|^2 +  |B^+_{\!-\frac12}|^2}{|B^-_{\frac12}|^2 + |B^-_{-\frac12}|^2}
\,.
\end{align}
They are an extended definitions from two-body decay, based on asymmetry parameters in terms of  partial wave amplitudes from Lee-Yang~\cite{Lee:1957qs} but in helicity formalism. The
parameter $\gammaXi$ represents the relative magnitude of the two types of degenerate amplitudes. 

The decay $\Xi_c^+ \to \Xi^- \pi^+ \pi^+$ may include intermediate resonance processes from three two-body decays: $\Xi_c^+ \to \Xi(1530)^0 \pi^+\to\Xi^-\pi^+\pi^+$, $\Xi_c^+ \to \Xi(1620)^0 \pi^+\to\Xi^-\pi^+\pi^+$, and $\Xi_c^+ \to \Xi(1690)^0 \pi^+\to\Xi^-\pi^+\pi^+$. Among these, the first process is suppressed due to the spin of $\Xi(1530)^0$ being $\frac{3}{2}$. The intermediate states $\Xi(1620)^0$ and $\Xi(1690)^0$ are allowed and can also be described by our method.
The amplitudes are model-independent and are relative to intermediate states by the integration:
\begin{align}
\label{eq:B-int}
|B^\mu_{\lambda_2}|^2
= \int d\Phi_3 
|  \tilde{B}^\mu_{\lambda_2}(m_{\Xi^-\pi^+},m_{\Xi^-\pi'^+},m_{\pi^+\pi'^+}) |^2,
\end{align}
where $m_{\Xi^-\pi^+}$ and $m_{\Xi^-\pi'^+}$ are the invariant mass of $\Xi^-$ with two pions respectively, also
are treated as Dalitz plots variables~\cite{JPAC:2019ufm} and the space $d\Phi_3$ consists of Euler angles given in Tab.~\ref{tab:decay}. The $m_{\pi^+\pi'^+}$ is invariant mass of two pions.
The integration can be estimated by using Monte-Carlo method but is out of this study, so we keep the amplitudes arbitrary for generality, and clarify the magnitude and phase angle of an amplitude by denoting $B^+_{\lambda_2} = b^+_{\lambda_2} e^{i\zeta_{\lambda_2}^+}$, then the four magnitudes can be rewritten in terms of the asymmetry parameters as
\begin{align}\label{eq:amp-b}
(b^+_\frac12)^2 &= 1
\,, \nonumber \\
(b^+_{-\frac12})^2 &= 
\frac{1-\alpha_{\Xi^+_c}}{1+\alpha_{\Xi^+_c}}
\,,\nonumber\\
(b^-_\frac12)^2 &= \frac{1+\betaXi}{\gammaXi(1+\alphaXi)}
\,, \nonumber\\
(b^-_{-\frac12})^2 &= 
\frac{1-\betaXi}{\gammaXi(1+\alphaXi)}.
\end{align}

\begin{figure}[thb]
	\begin{center}
\includegraphics[scale=.5]{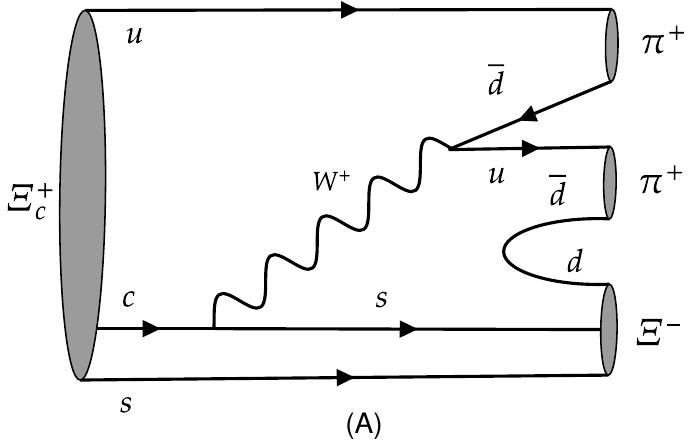}
\includegraphics[scale=.5]{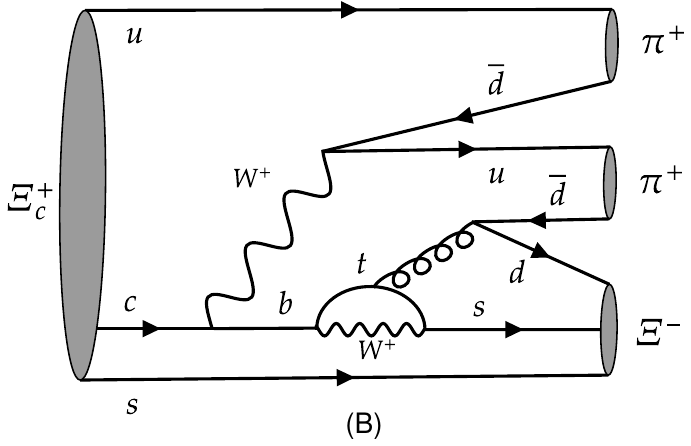}
	\end{center}\vspace{-5ex}
	\caption{Tree and penguin diagrams as examples for $\Xi_c^+ \to \Xi^- \pi^+\pi^+$ decay. }
	\label{fig:tree-loop}
\end{figure}

Analogous to \eq{abc}, the asymmetry parameters $\alpha_{\bar{\Xi}^-_c}$, $\beta_{\bar{\Xi}^-_c}$, and $\gamma_{\bar{\Xi}^-_c}$ for decay of the antibaryon $\bar{\Xi}^-_c\rightarrow \bar{\Xi}^+ \pi^-\pi^-$ have similar expressions. With those parity parameters, the $CP$ violation can be characterized by 
\begin{align}
\label{eq:ABC}
\mathcal{A}_{CP}
= &
\frac{ \alpha_{\Xi^+_c} + \alpha_{\bar{\Xi}^-_c}  }{\alpha_{\Xi^+_c} - \alpha_{\bar{\Xi}^-_c} }
\,,\nonumber \\
\mathcal{B}_{CP}
= &
\frac{ \beta_{\Xi^+_c} + \beta_{\bar{\Xi}^-_c}  }{\beta_{\Xi^+_c} - \beta_{\bar{\Xi}^-_c} }
\,,\nonumber \\
\mathcal{C}_{CP}
= &
\frac{ \gamma_{\Xi^+_c} - \gamma_{\bar{\Xi}^-_c}  }{\gamma_{\Xi^+_c} + \gamma_{\bar{\Xi}^-_c} }\,,
\end{align}
which have an advantage that the systematic uncertainties of production and detection asymmetries are largely cancelled \cite{Wang:2022tcm}. The parameters can be written in other expressions after inserting partial amplitudes. For example, the leading-power expansion of the first parameter is reduced to~\cite{BESIII:2021ypr,Donoghue:1986hh}
\begin{align}
\mathcal{A}_{CP}
\propto
-\tan\Delta\delta \tan\Delta\phi\,,
\end{align}
where the relative strong phase $\Delta\delta$ depends on final states, and the weak phase $\Delta\phi$ stems from interference between amplitudes with different partial wave configurations in the same decay~\cite{Saur:2020rgd}. The amplitudes partially derive from tree and penguin diagrams like the examples shown in Fig.~\ref{fig:tree-loop}, where the interfered CKM matrix elements $V^*_{cs}V_{ud}\times V^*_{cb}V_{tb}V^*_{ts}V_{ud}$ and their conjugations give rise to the nonzero phase $\Delta\phi$.

The tree diagram is Cabibbo-favored, while the penguin diagram is suppressed from off-diagonal elements $V^*_{cb}V^*_{ts}$. 
Following the measured values of CKM matrix~\cite{ParticleDataGroup:2020ssz}, the magnitude of the weak phase shift is about $\text{Im}[-(V^*_{cb}V_{tb}V^*_{ts}V_{ud})]$ and expected at order of  $\mathcal{O}(10^{-4}\!\sim\!10^{-5})$. Due to the possible suppression of the strong phase, the magnitude of $CP$ violation is expected to  be smaller than the expected value, and the interference effect between the two figures in Fig.~\ref{fig:tree-loop} also tends to suppress the observation of the violation. If an enhancement of $CP$ violation is observed, it arises from a combination of direct $CP$ violation and potential new physics contributions. To determine these effects, further measurements need to be performed at STCF in the future.

\section{Helicity system}\label{sec:helicity}
\begin{figure}[thb]
	\begin{center}
\includegraphics[scale=.4]{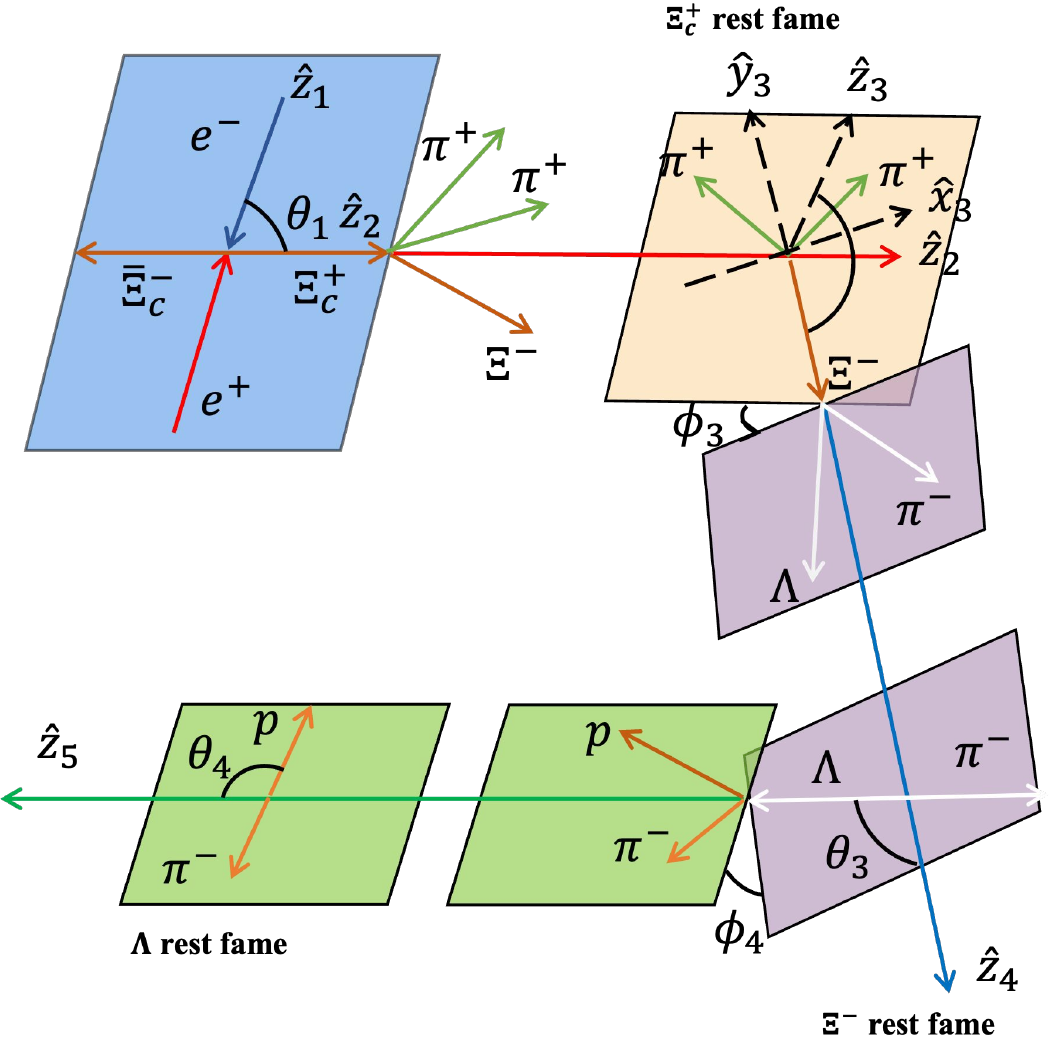}
	\end{center}\vspace{-5ex}
	\caption{Definition of helicity frame at $e^+e^-$ collider.}
	\label{fig:angle}
\end{figure}

In this analysis, we adopt a helicity frame to describe the decay chain~\cite{{0025}}. The helicity angles and amplitudes of the $\Xi_{c}^{+}\bar{\Xi}_{c}^-$ production and decay are listed in Tab.~\ref{tab:decay}, and the corresponding angles are also shown in Fig.~\ref{fig:angle}. The momentum $p_i$ are obtained by boosting particle $i$ to the rest frame of its mother particle. 
 \begin{table}[htbp]
\caption{Helicity angles and amplitudes in relative decays.} 
\label{tab:decay}
\begin{tabular}{ccc}
\hline
decay & helicity angle &helicity amplitude\\
$\gamma^*\rightarrow \Xi^+_c(\lambda_1)  \bar{\Xi}^-_c(\lambda_0)$ & $(\theta_1, \phi_1)$ & $A_{\lambda_1,\lambda_0}$   \\
$\Xi^+_c(\lambda_1)\rightarrow \Xi^-(\lambda_2)  \pi^+\pi^+$ & $(\phi_2,\theta_2,\psi_2)$ & $B_{ \lambda_2 }^\mu$   \\
$\Xi^-(\lambda_2)\rightarrow \Lambda(\lambda_3)  \pi^-$ & $(\theta_3, \phi_3)$ & $F_{ \lambda_3 }$   \\
$\Lambda(\lambda_3)\rightarrow p(\lambda_4)  \pi^-$ & $(\theta_4, \phi_4)$ & $H_{ \lambda_4 }$   \\
 \hline
\end{tabular}
\end{table}

The angle between the $\Xi^+_c$ production and the collision plane is defined as $\phi_1$. For the three-body decay of $\Xi^+_c \to \Xi^- \pi^+ \pi^+$, the Euler angles $(\phi_2,\theta_2,\psi_2)$ are used. The system is rotated from $\hat{z}_{2}$ to $\hat{z}_{3}$ following the ZYZ convention. In $\Xi^-$ decays, a rotation from $\hat{z}_{3}$ to $\hat{z}_{4}$ is needed. Due to the rotational invariance of helicity, this rotation operation does not introduce any additional modifications. 
In fact, the angle $\psi_2$ will not contribute to integrated cross sections we are interested in, and the selection of pion does not alter amplitudes since the meson's spin is zero.

\section{Angular distributions}
The spin and polarization information of particles can be encoded in spin density matrix (SDM)~\cite{Doncel:1973sg,Chen:2020pia}. The SDM of spin-$\frac12$ particles like $\Xi^+_c$ is expressed by a $2\times2$ matrix:
\begin{align}
\rho^{\Xi^+_c}
= &
\frac{\cP_0}{2}
\begin{pmatrix}
1 + \cP_z & \cP_x - i\cP_y \nonumber\\
\cP_x + i\cP_y & 1 - \cP_z
\end{pmatrix}\,,
\end{align}
where $\cP_0$ carry the unpolarized information. Polarization $\cP_z$ is longitudinal part, while the information of transverse polarization is taken by $\cP_x$ and $\cP_y$. To obtain their explicit expressions, transfer functions starting from initial collision will be solved in the following sections.

\subsection{$\Xi^+_c$ production on $e^+e^-$ collider}
The SDM of $\Xi^+_c$ from polarized virtual photon can be written as
\begin{eqnarray}\label{eq:Xi-rho}
\rho^{\Xi^+_c}_{\lambda_1,\lambda_1^{'}}&\propto&\sum_{{\lambda},\lambda_0}{\rho^{\gamma^*}_{\lambda,\lambda^{'}}}D^{1*}_{\lambda,\lambda_{1}-\lambda_0}(\phi_1,\theta_1,0)\nonumber\\
	&\times&D^{{1}}_{\lambda',\lambda_{1}^{'}-\lambda_0}(\phi_1,\theta_1,0)A_{\lambda_{1},\lambda_0}A_{\lambda_{1}^{'},\lambda_0}^{*},
\end{eqnarray}
where $D^J_{\lambda_j,\lambda_k}$ is the Wigner-D function. For polarized symmetric $e^+e^-$ beams, if the transverse polarization information is considered, the SDM of photon is given by~\cite{Cao:2024tvz}
\begin{eqnarray}
\rho^{\gamma^*}_{\lambda_0,\lambda_0^{'}}
= &
\frac12  \begin{pmatrix}
 1 & 0 & p_T^2\\
 0 & 0 & 0 \\
 p_T^2 & 0 & 1
\end{pmatrix}\,,
\end{eqnarray}
where the range of transverse polarization $p_T$ satisfies $0<p_T<1$. Carrying out \eq{Xi-rho} under parity conversation, we get the unpolarized cross section which depends on the polarization of virtual photon, that is 
\begin{align}
\label{eq:P0-photon}
\cP_0 = 1 + \alpha_c\cos^2\theta_1 + p_T^2 \,\alpha_c  \sin^2\theta_1
\cos2\phi_1\,.
\end{align}
The constant $\tfrac12|A_{\frac12,-\frac12}|^2+|A_{\frac12,\frac12}|^2$ is suppressed, since it does not contribute to the final normalized cross sections. The angular distribution parameter $\alpha_c$ for $\ee\rightarrow \Xi^+_c\bar{\Xi}^-_c$ is defined by
\begin{align}
	\alpha_c={|A_{\half,-\half}|^2-2|A_{\half,\half}|^2\over |A_{\half,-\half}|^2+2|A_{\half,\half}|^2}\,.
\end{align}
The transverse and longitudinal  polarizations of $\Xi^+_c$ are 
\begin{align}\label{eq:Px-Py}
\cP_x = &
\frac{-p_T^2\sqrt{1-\alpha_c^2}\sin\Delta_1\sin\theta_1\sin2\phi_1 }{1 + \alpha_c\cos^2\theta_1 + p_T^2 \,\alpha_c  \sin^2\theta_1
\cos2\phi_1},
\nonumber\\
\cP_y = &
\frac{\sqrt{1-\alpha_c^2}\sin\Delta_1\sin\theta_1\cos\theta_1(1-p_T^2\cos2\phi_1) }{1 + \alpha_c\cos^2\theta_1 + p_T^2 \,\alpha_c  \sin^2\theta_1
\cos2\phi_1}
,\nonumber\\
\cP_z = & 
0,
\end{align}
where $\Delta_1$ represents phase angle difference $\Delta_1 = \zeta_{\frac12,-\frac12}-\zeta_{\frac12,\frac12}$. 
In symmetric electron-positron collision experiments, the transverse polarization of the positron and electron beams directly influences the transverse polarization of $\Xi^+_c$ as well as $\bar{\Xi}^-_c$. 
Though the measured values of the parity parameter of this process is lack, in the following sections we use $\alpha_c = 0.7$ and $\Delta_1=\pi/6$ which are close to the measurements of strange baryons~\cite{BESIII:2016nix}.

\subsection{$\Xi^+_c(\lambda_1)\rightarrow \Xi^-(\lambda_2)  \pi^+\pi^+ $ }
The information of parity violation of this three-body decay is carried by the SDM of $\Xi^-$, which is given by transfer equation:
\begin{eqnarray}\label{eq:SDM-spin12}
\rho^{\Xi^-}_{\lambda_{2},\lambda'_{2}}&\propto&\sum_{{\lambda_{1}},\lambda'_{1}, \mu}{\rho^{\Xi^+_c}_{\lambda_{1},\lambda_{1}^{'}}}D^{\frac{1}{2}*}_{\lambda_{1},\mu}(\phi_2,\theta_2,\psi_2)\nonumber\\
	&\times&D^{\frac{1}{2}}_{\lambda_{1}^{'},\mu}(\phi_2,\theta_2,\psi_2)B^\mu_{\lambda_{2}} B_{\lambda_{2}^{'}}^{\mu*}\,,
\end{eqnarray}
where $\mu$ is the z-component of angular momentum of $\Xi^+_c$ whose quantization axis itself rotates under the rotation of the system.
After summing all combinations and implementing  simplifications, the polarization operators of $\Xi^-$ reduce to 
\begin{align}
\cP_0^{\Xi^-}
= &
\frac{\cP_0}{2}
\left(
f^{\Xi^-}_{+}
+
f^{\Xi^-}_{-}\cP_{xy}\sin\theta_2
\right)
\,,\nonumber \\
\cP_0^{\Xi^-}\cP_z^{\Xi^-}
= & 
\frac{\cP_0}{2}\left(
g^{\Xi^-}_{+}
+
g^{\Xi^-}_{-}\cP_{xy}\sin\theta_2
\right)
\,, \nonumber\\
\cP_0^{\Xi^-}\cP_x^{\Xi^-}
= &
\cP_0\bigg[
\cos\Delta\!^+\left(
1+\cP_{xy}\sin\theta_2
\right)b^+_\frac12 b^+_{-\frac12}
\nonumber \\
&\quad +
\cos\Delta\!^-\left(
1-\cP_{xy}\sin\theta_2
\right)b^-_\frac12 b^-_{-\frac12}
\bigg]
\,, \nonumber\\
\cP_0^{\Xi^-}\cP_y^{\Xi^-}
= &
-\cP_0\bigg[
\sin\Delta\!^+\left(
1+\cP_{xy}\sin\theta_2
\right)b^+_\frac12 b^+_{-\frac12}
\nonumber \\
&\qquad +
\sin\Delta\!^-\left(
1-\cP_{xy}\sin\theta_2
\right)b^-_\frac12 b^-_{-\frac12}
\bigg],
\label{eq:P-Xim}
\end{align}
where the functions $f^{\Xi^-}_{\pm}$ and $g^{\Xi^-}_{\pm}$ carry the dynamical information of the decay via \eq{amp-b} and are defined by
\begin{align}
f^{\Xi^-}_{\pm}
= & (b^+_{\frac12})^2 + (b^+_{-\frac12})^2 
\pm
\left[
(b^-_{\frac12})^2 + (b^-_{-\frac12})^2
\right]
\,,\nonumber\\
g^{\Xi^-}_{\pm}
= & (b^+_{\frac12})^2 - (b^+_{-\frac12})^2 
\pm
\left[
(b^-_{\frac12})^2 - (b^-_{-\frac12})^2
\right]\,.
\end{align}
Especially, there are $g^{\Xi^-}_\pm=0$ when parity is conserved.
The transverse information from $\Xi^+_c$ is involved via the definition  $\cP_{xy}=\cP_x\cos\phi_2+\cP_y\sin\phi_2$. The transverse polarizations of $\Xi^-$ depend on phase differences which are defined by $\Delta_{\pm} = \zeta^{-}_{\pm\frac12} - \zeta^{+}_{\pm\frac12}$ and $\Delta\!^{\pm}=\zeta^{\pm}_{-\frac12}-\zeta^{\pm}_\frac12$ that lead to a  relation $\Delta_+-\Delta_-+\Delta^--\Delta^+=0$. There are also other definitions of parity parameters for three-body system adopted in $\Lambda^+_c\rightarrow p K^- \pi^+$ decay~\cite{Wei:2022kem}, and are relative to our parameters by
\begin{align}
\mathcal{G}_0 = &
\frac{\gamma^{\Xi^+_c}-1}{\gamma^{\Xi^+_c}+1}
\,,\nonumber\\
\mathcal{G}_1 = &
\kappa_-\sin\Delta_- +
\kappa_+\sin\Delta_+
\,,\nonumber\\
\mathcal{G}_2 = &
\kappa_-\cos\Delta_-
+
\kappa_+\cos\Delta_+\,,
\label{eq:cG-012}
\end{align}
where 
\begin{align}
\kappa_{\pm}= \frac{\sqrt{(1\pm\alphaXi)(1\pm\betaXi)\gammaXi}}{1+\gammaXi}\,.
\end{align}
When parity is conserved, the values of  parameters $\mathcal{G}_1$ and $\mathcal{G}_2$ can not be determined directly due to their undetermined phases in \eq{cG-012}, while $\alphaXi=\betaXi=0$ that provides more freedoms for cross testing of parity violations.

\subsection{Cascade decays and joint angular distributions}
\begin{figure}[thb]
	\begin{center}
	\includegraphics[scale=.50]{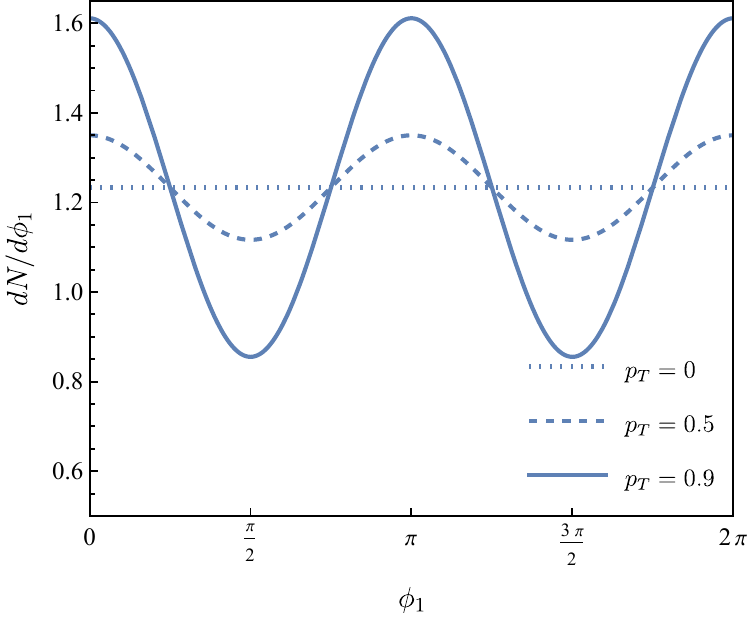}
	\end{center}\vspace{-5ex}
	\caption{Angular distributions of $\phi_1$ for $p_T=0$, $0.5$, $0.9$ (dotted, dashed and solid lines, respectively).  }
	\label{fig:N-theta1}
\end{figure}

\begin{figure}[thb]
	\begin{center}
		\includegraphics[scale=.50]{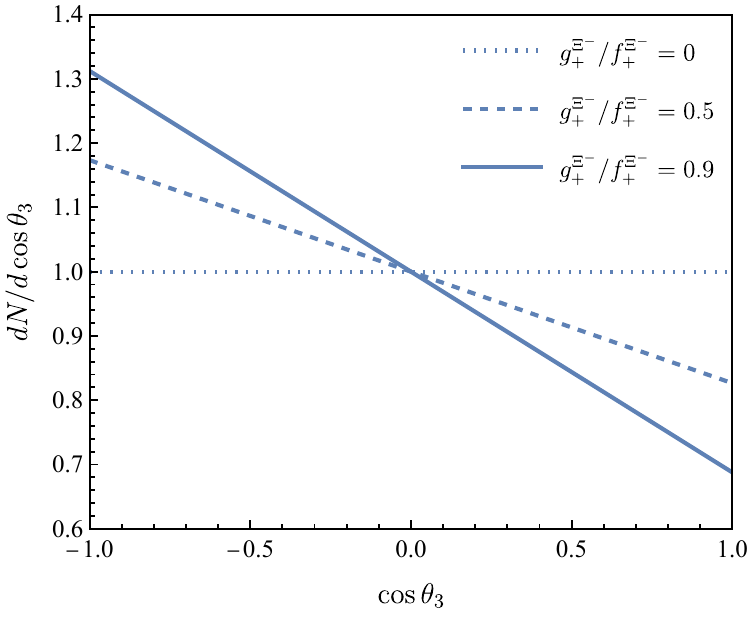}
	\end{center}\vspace{-5ex}
	\caption{Angular distributions of $\cos\theta_3$ for $g^{\Xi^-}_+/f^{\Xi^-}_+=0$, $0.5$, $0.9$ (dotted, dashed and solid lines, respectively).  }
	\label{fig:N-theta3}
\end{figure}

The dominate processes for $\Xi^-$ decay are $ \Xi^-\rightarrow \Lambda  \pi^- $ and $ \Lambda\rightarrow p  \pi^- $. Their helicity angles and amplitudes are defined in Sec.~\ref{sec:helicity}, and the asymmetry parameter of the spin-$\frac12$ baryons for the  decay $ \Xi^-\rightarrow \Lambda  \pi^- $ can be defined as
\begin{align}
\alpha_{\Xi^-} = 
{|F_{\frac12}|^2-|F_{-\frac12}|^2\over |F_{\frac12}|^2+|F_{-\frac12  }|^2}\,.
\end{align}
 The parameter $\alpha_\Lambda$ for $ \Lambda\rightarrow p  \pi^- $ has a similar expression. The latest measurements of the two parameters are $\alpha_{\Xi^-}=-0.367^{+0.005}_{-0.006}$ and $\alpha_{\Lambda}=0.746\pm 0.008$ from the Particle Data Group (PDG)~\cite{BESIII:2023jhj}. Carrying out an equation analog to \eq{SDM-spin12}, the unpolarized and longitudinal parts of $\Lambda$ can be obtained, they are
\begin{align}
\label{eq:P0-Xim}
\cP_0^{\Lambda}
= &
\frac{2\cP_0^{\Xi^-}}{1+\alpha_{\Xi^-}}
\bigg[
1 +\alpha_{\Xi^-}
\cos\theta_3 \cP^{\Xi^-}_z
\nonumber \\
& \qquad\quad +
\alpha_{\Xi^-}\sin\theta_3 
\left(\cos\phi_3\cP_x^{\Xi^-}  
+ 
\sin\phi_3\cP_y^{\Xi^-} 
\right)
\bigg]\,,
 \\
\cP_0^{\Lambda}\cP_z^{\Lambda}
= &
\frac{2\cP_0^{\Xi^-}}{1+\alpha_{\Xi^-}}
\bigg\{
\alpha_{\Xi^-}+
\cos\theta_3 \cP^{\Xi^-}_z
\nonumber \\
& \qquad\qquad +
\sin\theta_3 
\left(\cos\phi_3\cP_x^{\Xi^-}  
+ 
\sin\phi_3\cP_y^{\Xi^-} 
\right)
\bigg\}\,.
\end{align}
The SDM of proton can be obtained in a same way, 
and the angular distribution is given by $\mathcal{W}=\text{Tr}\rho^p$, which is similar to \eq{P0-Xim} but in terms of polarizations of  $\Lambda$. Next the distributions for polar angles can be obtained by integrating out other angles, and the nontrivial distributions are
\begin{align}
\label{eq:N-theta1}
\frac{dN}{d\phi_1}
\propto &
1 + \alpha_c\frac{1+2 p^2_T}{3}\cos2\phi_1
\,, \\
\frac{dN}{d\cos\theta_1}
\propto &
1 + \alpha_c\cos^2\theta_1 
\,, \\
\label{eq:N-theta3}
\frac{dN}{d\cos\theta_3}
\propto &
1 + \alpha_{\Xi^-}\frac{g^{\Xi^-}_+}{f^{\Xi^-}_+}\cos\theta_3
\,, \\
\label{eq:N-theta4}
\frac{dN}{d\cos\theta_4}
\propto &
1 + \alpha_{\Xi^-}\alpha_{\Lambda}
\cos\theta_4
\,,
\end{align}
where the distribution of $\cos\theta_1$ is free of beam polarization, and the last distribution \eq{N-theta4} is known in Ref.~\cite{BESIII:2023jhj} while the distribution of $\cos\theta_2$ is flat. 
The distribution of $\cos\theta_3$ is effected by parity parameters of $\Xi^+_c$ through $g^{\Xi^-}_+/f^{\Xi^-}_+=(\beta_{\Xi^+_c}+\alpha_{\Xi^+_c}\gamma_{\Xi^+_c})/(1+\gamma_{\Xi^+_c})$. 
The plots for \eq{N-theta1} and \eq{N-theta3} are given in Fig.~\ref{fig:N-theta1} and Fig.~\ref{fig:N-theta3}   respectively. Under the assumption of parity conversation  that $\alpha_{\Xi^-}=0$ for $\Xi^-\rightarrow \Lambda \pi^-$ or
$g^{\Xi^-}_+=0$ for $ \Xi^+_c(\lambda_1)\rightarrow \Xi^-(\lambda_2)  \pi^+\pi^+ $, the distributions turn to be constants respectively. 
Otherwise when parity violation exists, the absolute amplitudes of the distributions for $\cos\theta_3$ are non-zero.

The angular distribution after  normalization is defined by
\begin{align}
\widetilde{\cW}
= &
\frac{\cW(\theta_i,\phi_i,\alpha_i)}{\int \Pi^4_i d\cos\theta_i
\phi_i \cW(\theta_i,\phi_i,\alpha_i)}\,.
\end{align}

\begin{figure}[thb]
	\begin{center}
	\includegraphics[scale=.50]{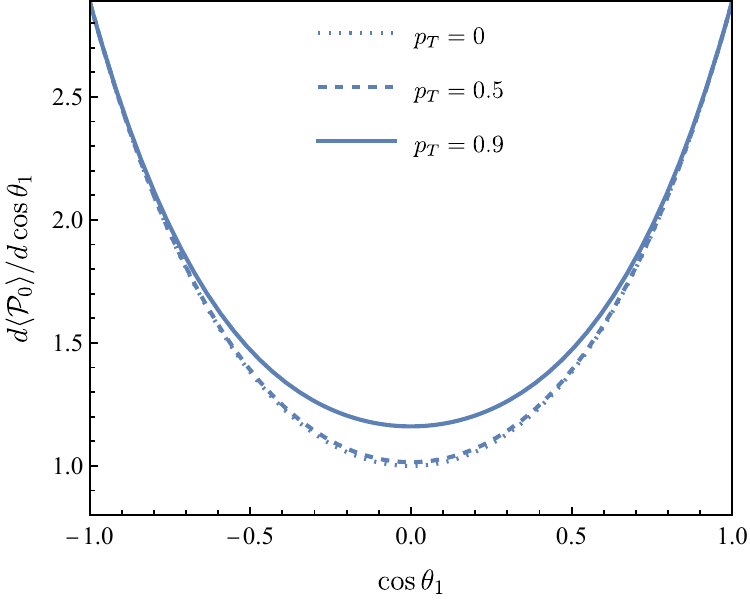}
	\end{center}\vspace{-5ex}
	\caption{Weighted $\cP_0$ as a function of $\cos\theta_1$ for $p_T=0$, $0.5$, $0.9$ (dotted, dashed and solid lines, respectively). }
	\label{fig:ave-P0}
\end{figure}

\begin{figure}[thb]
	\begin{center}
	\includegraphics[scale=.50]{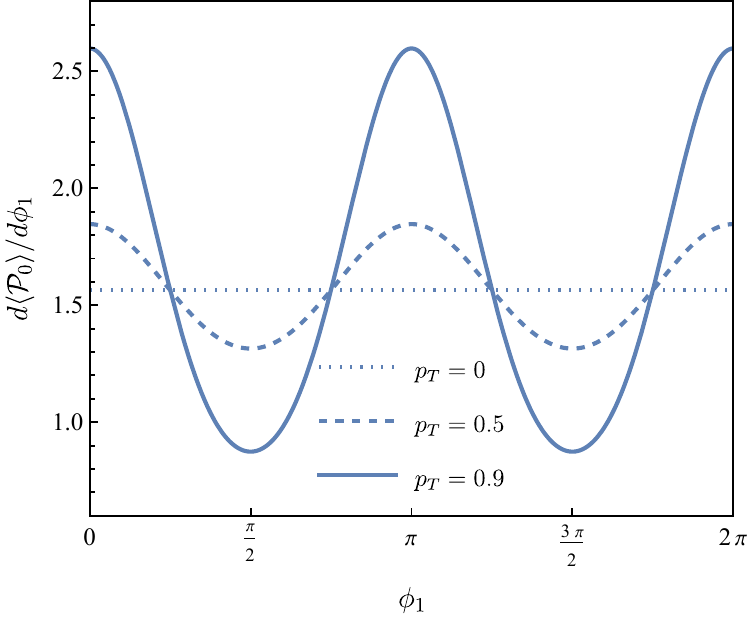}
	\end{center}\vspace{-5ex}
	\caption{Weighted $\cP_0$ as a function of $\phi_1$ for $p_T=0$, $0.5$, $0.9$ (dotted, dashed and solid lines, respectively). }
	\label{fig:ave-P0-2}
\end{figure}

\begin{figure}[thb]
	\begin{center}
	\includegraphics[scale=.50]{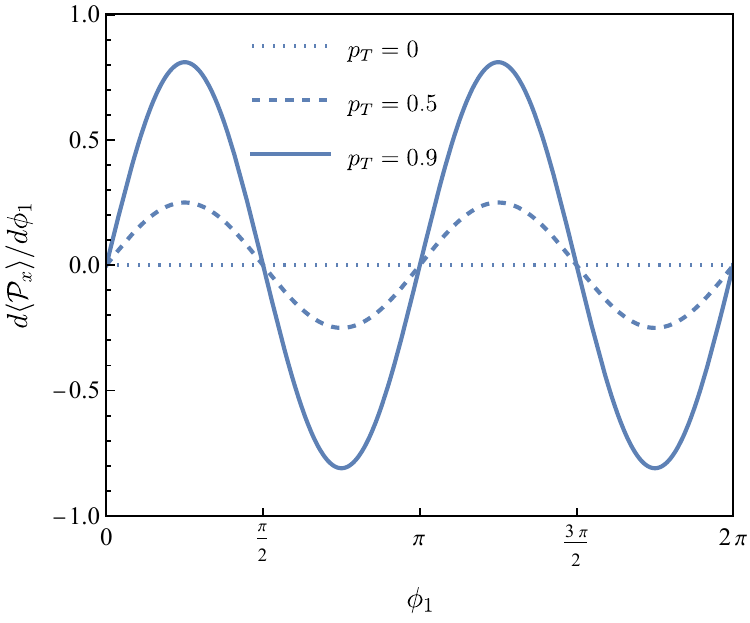}
	\end{center}\vspace{-5ex}
	\caption{Weighted $\cP_x$ as a function of $\phi_1$ for $p_T=0$, $0.5$, $0.9$ (dotted, dashed and solid lines, respectively). }
	\label{fig:ave-Px}
\end{figure}

\begin{figure}[thb]
	\begin{center}
	\includegraphics[scale=.50]{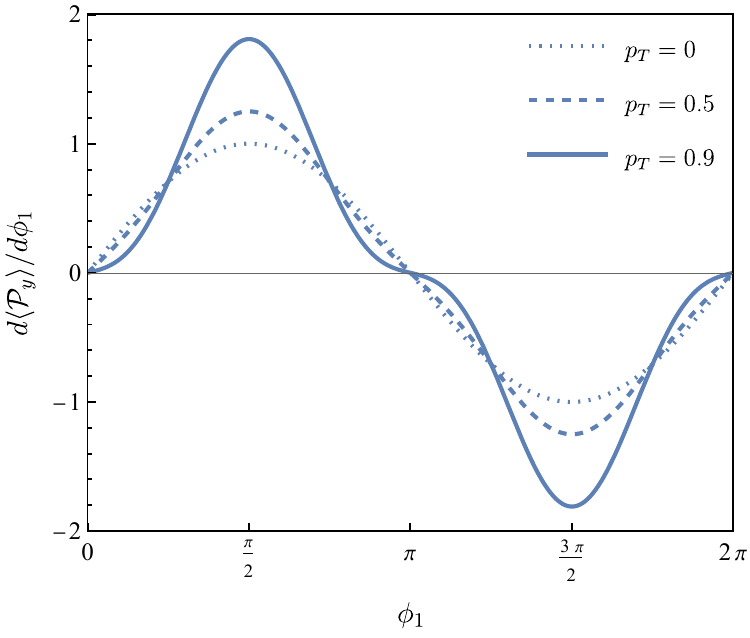}
	\end{center}\vspace{-5ex}
	\caption{Weighted $\cP_y$ as a function of $\phi_1$ for $p_T=0$, $0.5$, $0.9$ (dotted, dashed and solid lines, respectively). }
	\label{fig:ave-Py}
\end{figure}

It is directive to define the weighted polarizations of $\Xi^+_c$ depending on $\theta_1$ or $\phi_1$ by integrating out other angles to get 
\begin{align}
\frac{d\langle \cP_0 \rangle}{d\cos\theta_1}
\propto &\,
1 + 2\alpha_c\cos^2\theta_1
+
\alpha_c^2 \left(
\cos^4\theta_1 + \frac{p^4_T}{2} \sin^4\theta_1
\right)
\,, \\ 
\frac{d\langle \cP_0 \rangle}{d\phi_1}
\propto &
1 + \frac23\alpha_c(1+2p^2_T\cos2\phi_1)
\nonumber\\ &+
\frac{\alpha_c^2}{5}
\left(
1 + \frac43p^2_T\cos2\phi_1 +
\frac83p^4_T \cos^2(2\phi_1)
\right)
\,,\\
\frac{d\langle \cP_y \rangle}{d\phi_1}
\propto &\,
(1 - p_T^2\cos2\phi_1 ) \sin\phi_1
\,,\\
\frac{d\langle \cP_x \rangle}{d\phi_1}
\propto & \,
 p_T^2\sin2\phi_1\,,
\end{align}
and $d\langle \cP_z \rangle / d\phi_1
=  0$. The corresponding plots are shown in Fig.~\ref{fig:ave-P0}, ~\ref{fig:ave-P0-2}, ~\ref{fig:ave-Px} and ~\ref{fig:ave-Py} respectively. The beam polarization enhances the polarizations distributions of $\phi_1$, but suppress the unpolarization distribution of $\cos\theta_1$. 

For the observed data sample of $N$ events, the likelihood function is expressed as~\cite{Han:2019axh}
\begin{align}
L(\theta_i, \phi_i,\alpha_c,\alpha_{\Xi^-},\alpha_{\Lambda},\alpha_{\Xi^+_c},\beta_{\Xi^+_c},\gamma_{\Xi^+_c})
=
\prod^N_{i=1} \widetilde{\cW}_j\,,
\end{align}
where $\theta_i$ and $\phi_i$ represent polar angle and azimuth angle, and $\alpha_c$, $\alpha_{\Xi^-}$, $\alpha_{\Lambda}$, $\alpha_{\Xi^+_c}$, $\beta_{\Xi^+_c}$, and $\gamma_{\Xi^+_c}$ mean decay parameters, and the product is computed based on the probability of the $i$-th event $\widetilde{\cW}_j$. Here, $\alpha_{\Xi^-}$ and $\alpha_{\Lambda}$ are fixed to PDG values~\cite{BESIII:2023jhj}. According to the maximum likelihood method, the relative uncertainty for estimating statistical sensitivity to parity parameter $\alpha_{\Xi^+_c}$ is defined as
\begin{align} 
\delta(\alpha_{\Xi^+_c})
=
\frac{\sqrt{V(\alpha_{\Xi^+_c})}}{|\alpha_{\Xi^+_c}|} \,,
\end{align}
where the inverse of the variance is given by 
\begin{align}
V^{-1}(\alpha_{\Xi^+_c})
=
N\int \frac{1}{\widetilde{\cW}}
\left[ 
\frac{\partial \widetilde{\cW}}{\partial \alpha_{\Xi^+_c}}
\right]^2
\prod_{i}^4d\cos\theta_i d\phi_i\,.
\end{align}

Indeed for identifying significance of $CP$ violations, the statistical sensitivity of $\mathcal{A}_{CP}$ in \eq{ABC} can be estimated if $\alpha_{\Xi^+_c}$ and $\bar\alpha_{\Xi^-_c}$ are considered as non-correlation, via error propagation formula:
\begin{align}
\label{eq:delta-Acp}
\delta(\mathcal{A}_{CP})
= 
\frac{2\sqrt{\alpha_{\Xi^+_c}^2\delta(\bar{\alpha}_{\bar{\Xi}^-_c})^2 + \bar{\alpha}_{\bar{\Xi}^-_c}^2 \delta(\alpha_{\Xi^+_c})^2 } }{(\alpha_{\Xi^+_c} - \bar{\alpha}_{\bar{\Xi}^-_c})^2}\,,
\end{align}
where $\delta(\bar{\alpha}_{\bar{\Xi}^-_c})$ is the sensitivity for $\bar{\Xi}^-_c$ system. The sensitivities $\delta(\mathcal{B}_{CP})$ and $\delta(\mathcal{C}_{CP})$ for other two $CP$ parameters can be defined in similar ways.

\section{Numerical sensitivities}

\begin{figure}[H]
\begin{center}
\begin{overpic}[width=0.35\textwidth, trim=80 100 50 20,angle=0]{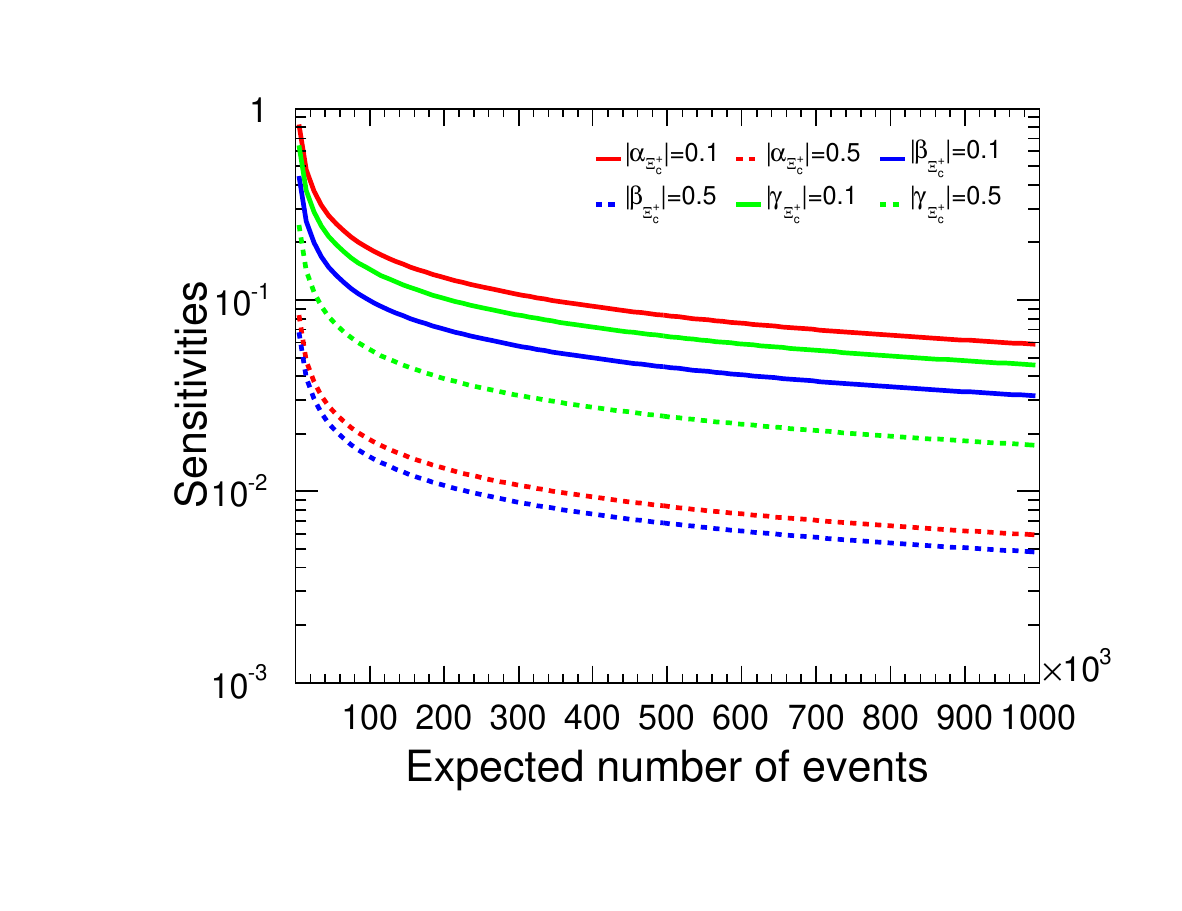}
\end{overpic}
\end{center}
\caption{The $\alpha_{\Xi_c^+}$, $\beta_{\Xi_c^+}$, and $\gamma_{\Xi_c^+}$ sensitivity distributions relative to signal yields in terms of different parameters.
The red solid and dashed lines represent the $\alpha_{\Xi_c^+}$ values of 0.1 and 0.5, respectively. The blue solid and dashed lines represent the $\beta_{\Xi_c^+}$ values of 0.1 and 0.5, respectively. The green solid and dashed lines represent the $\gamma_{\Xi_c^+}$ values of 0.1 and 0.5, respectively.
}
\label{fig:delta-alpha}
\end{figure}

We are particularly interested in the dependence of statistical sensitivities on asymmetry parameters and beam polarizations. To clarify the analysis, we fix $\alpha_c = 0.7$, $\Delta_1 = \pi/6$, and $\Delta_\pm=\Delta^+=\pi/3$. We have verified that other sets of phase angle values do not significantly affect the statistical quantities. In the following figures, when varying one parameter, the other two parameters are held constant. For example, the red lines in Fig.~\ref{fig:delta-alpha} illustrate the distributions of $\delta(\alpha_{\Xi^+_c})$ as functions of the event number $N$, with $\beta_{\Xi^+_c} = 0.1$ and $\gamma_{\Xi^+_c} = 0.5$ fixed. Similarly, the blue lines represent the distributions of $\delta(\beta_{\Xi^+_c})$ with $\alpha_{\Xi^+_c} = 0.1$ and $\gamma_{\Xi^+_c} = 0.5$, while the green lines show the distributions of $\delta(\gamma_{\Xi^+_c})$ with $\alpha_{\Xi^+_c} = 0.1$ and $\beta_{\Xi^+_c} = 0.3$. These sensitivities exhibit a negative correlation with the absolute values of the parameters. Under our chosen parameter values at $\alphaXi=0.5$ and  $\betaXi=0.5$, the corresponding  sensitivities reach $1\%$ with about $3\times10^3$ signal events.

\begin{figure}[H]
\begin{center}
\begin{overpic}[width=0.35\textwidth, trim=80 100 50 20,angle=0]{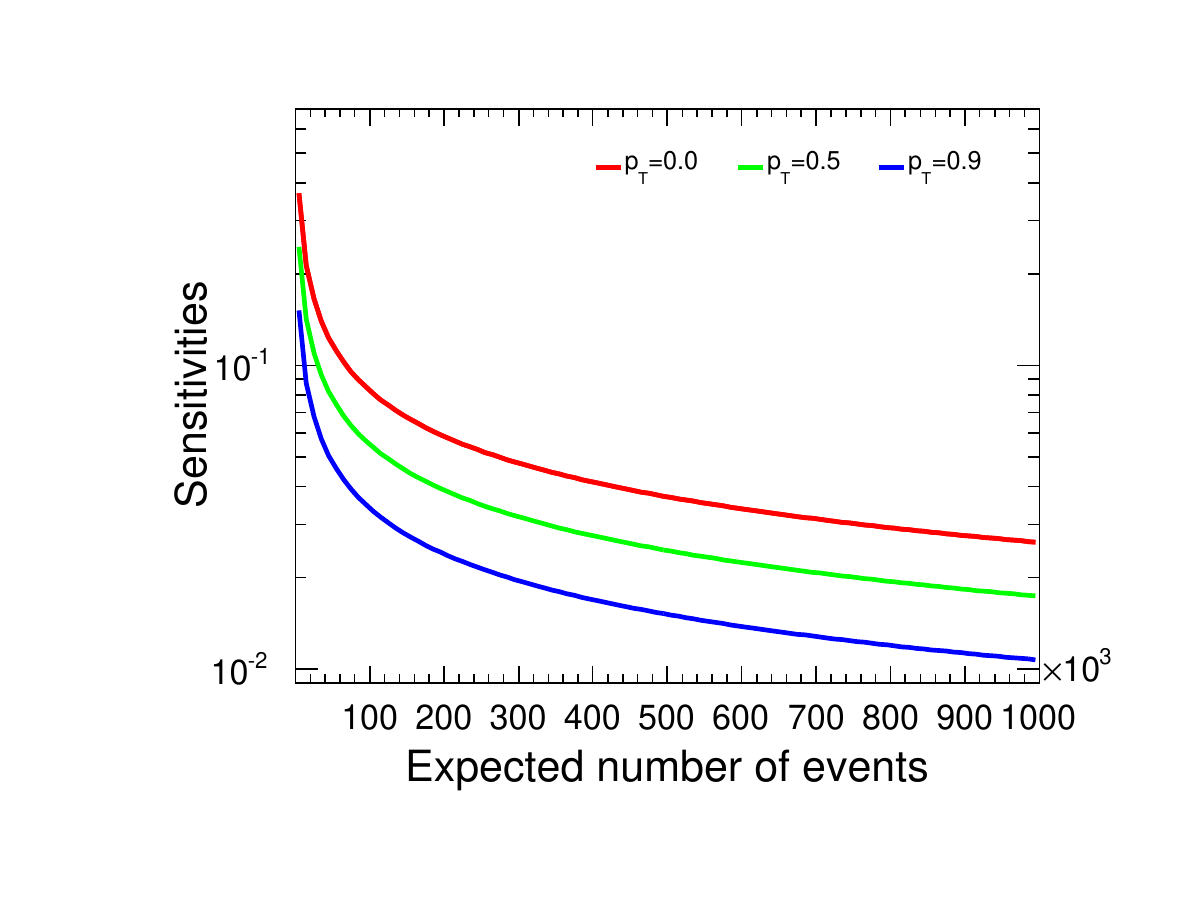}
\end{overpic}
\end{center}
\caption{The $\gamma_{\Xi_c^+}$ sensitivity distribution relative to signal yields assuming $\gammaXi=0.5$.
The $p_{T}$ values of 0.0, 0.5, and 0.9 correspond to the red, green, and blue lines, respectively.}
\label{fig:delta-alpha-Pt}
\end{figure}

\begin{figure}[H]
\begin{center}
\begin{overpic}[width=0.35\textwidth, trim=80 100 50 20,angle=0]{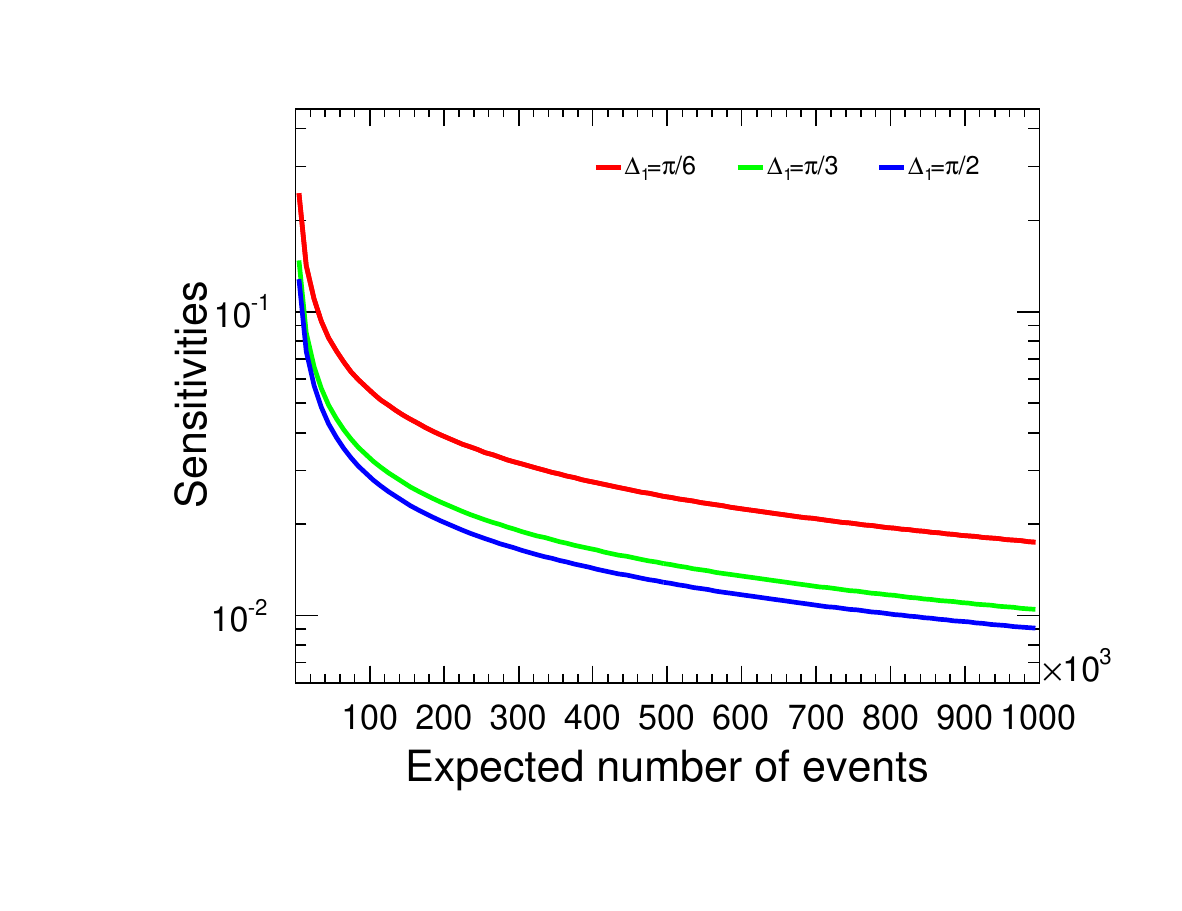}
\end{overpic}
\end{center}
\caption{The $\gamma_{\Xi_c^+}$ sensitivity distributions relative to signal yields assuming $\gammaXi=0.5$.
The $\Delta_{1}$ values of $\pi/6$, $\pi/3$, and $\pi/2$ correspond to the red, green, and blue lines, respectively.
}
\label{fig:delta-alpha-Delta}
\end{figure}

In Fig.~\ref{fig:delta-alpha}, The statistical significance of $\gamma_{\Xi^+_c}$ is weak respected to $\alpha_{\Xi^+_c}$ and $\beta_{\Xi^+_c}$ under same signal events. However, $\delta(\gammaXi)$ has strong dependence on beam polarizations since its parameter is defined as the radio of two projections of angular momentum in \eq{abc}. As exhibited in Fig.~\ref{fig:delta-alpha-Pt}, the measurement accuracy of $\gammaXi$ can be improved due to the beam polarization contribution. The significance of $\gammaXi$ is also relative to intrinsic phases difference $\Delta_1$ according to \eq{Px-Py}. If $\Delta_1$ reach to saddle point $\pi/2$, the requirement of events decreases apparently and the corresponding curves are plotted in Fig.~\ref{fig:delta-alpha-Delta}. In addition, we have test that $\alphaXi$ and $\betaXi$ sensitivities are not sensitive to $p_T$ and $\Delta_1$. 

\begin{figure}[H]
\begin{center}
\begin{overpic}[width=0.35\textwidth, trim=80 100 50 30,angle=0]{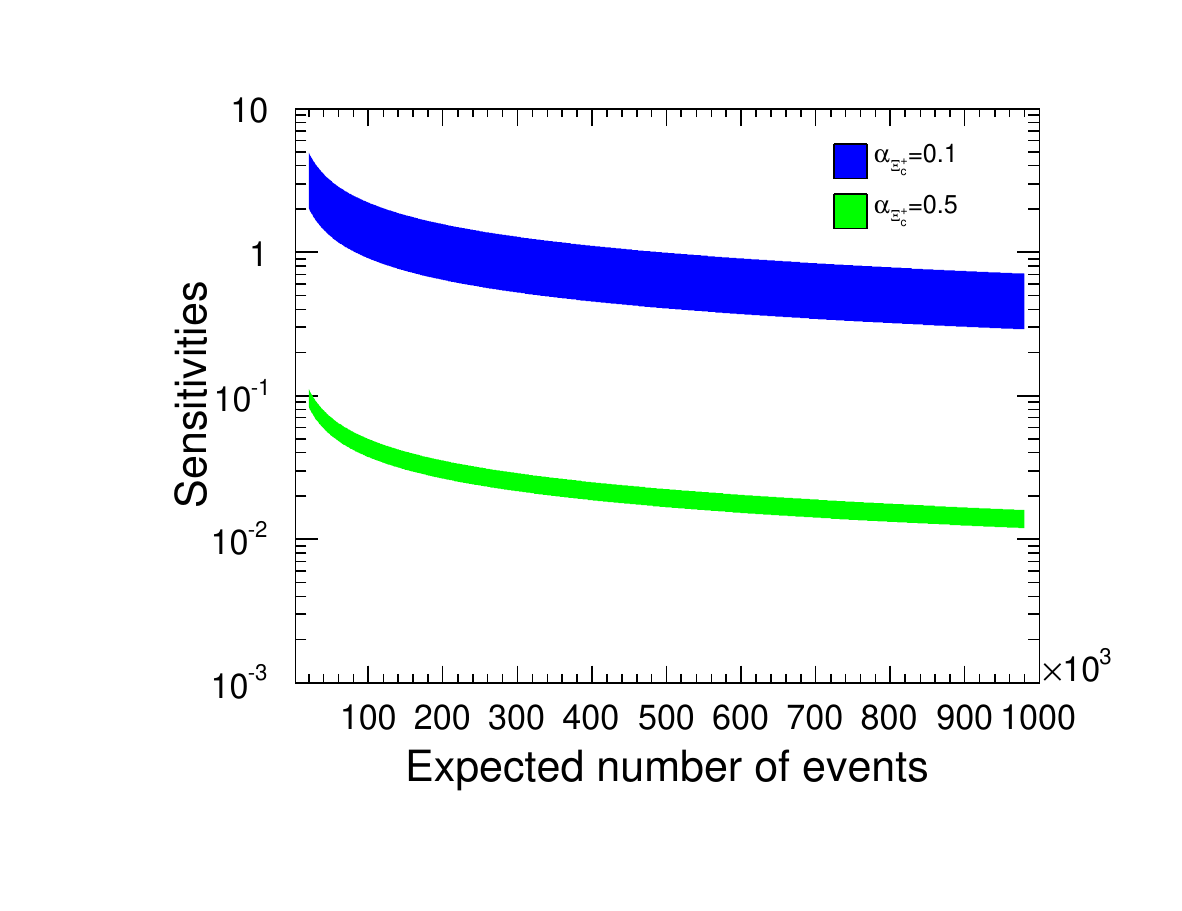}
\end{overpic}
\end{center}
\caption{The $\mathcal{A}_{CP}$ sensitivity distributions relative to signal yields.
The $\alpha_{\Xi^+_c}$ values of $0.1$ and $0.5$ correspond to the blue and green bands, respectively.
} 
\label{fig:alpha-cp}
\end{figure}

\begin{figure}[H]
\begin{center}
\begin{overpic}[width=0.35\textwidth, trim=80 100 50 20,angle=0]{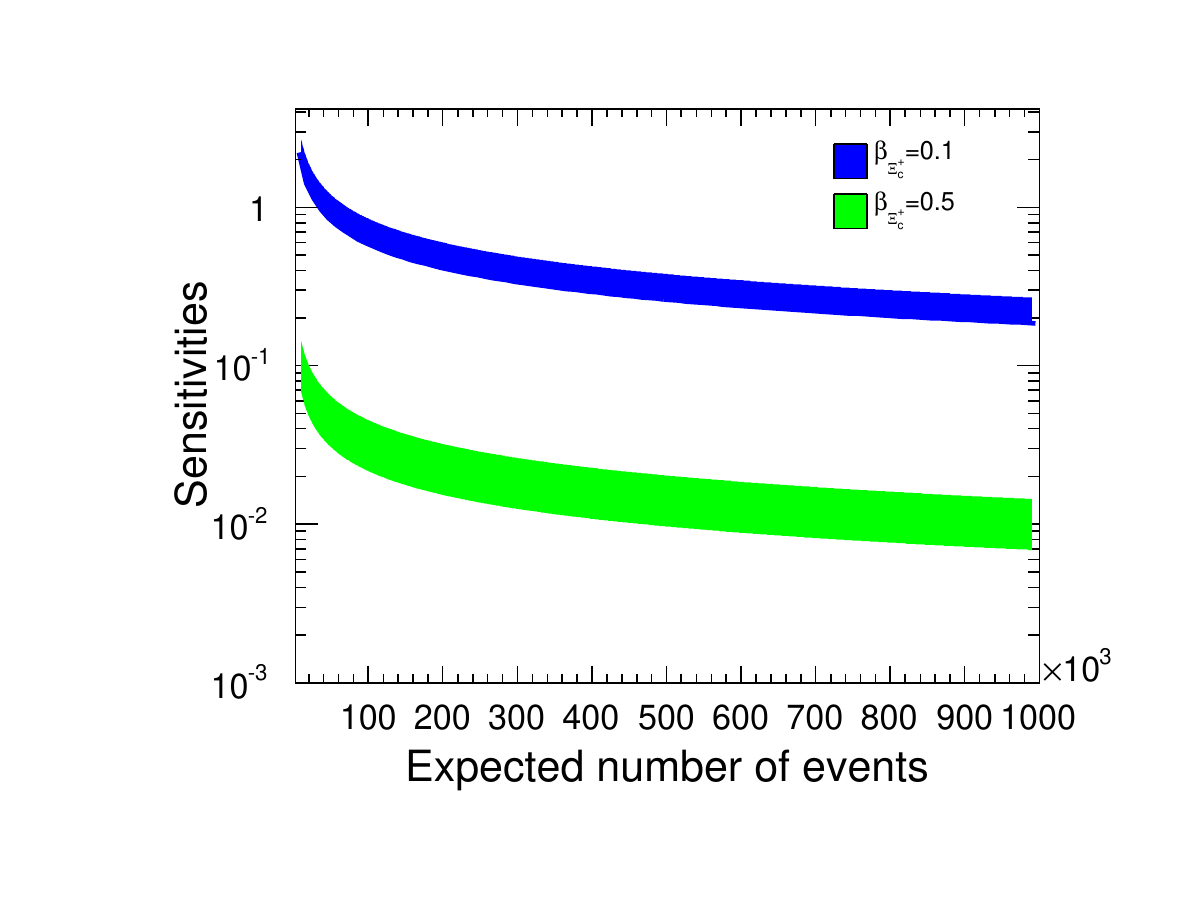}
\end{overpic}
\end{center}
\caption{The $\mathcal{B}_{CP}$ sensitivity distributions relative to signal yields. The $\beta_{\Xi^+_c}$ values of $0.1$ and $0.5$ correspond to the blue and green bands, respectively.}
\label{fig:beta-cp}
\end{figure}

\begin{figure}[H]
\begin{center}
\begin{overpic}[width=0.35\textwidth, trim=80 100 50 20,angle=0]{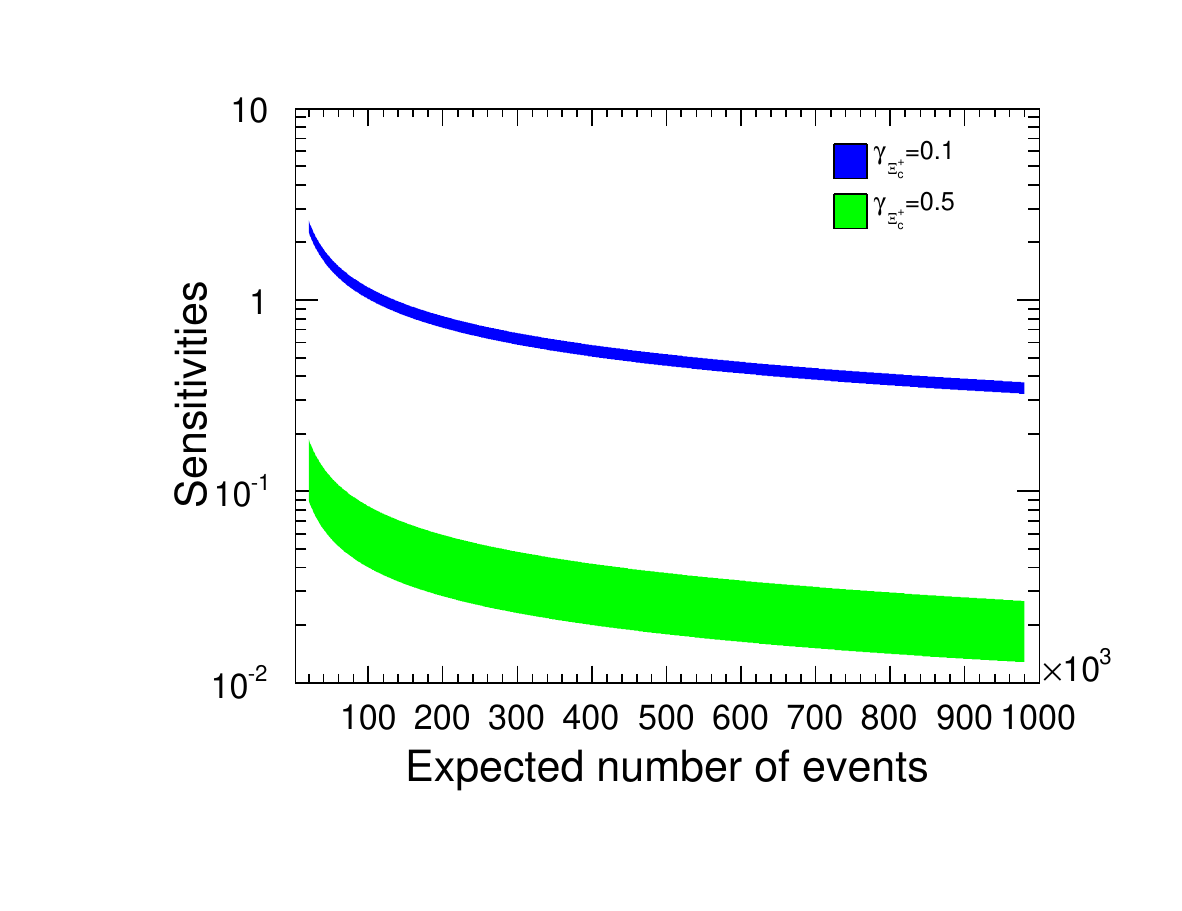}
\end{overpic}
\end{center}
\caption{The $\mathcal{C}_{CP}$ sensitivity distributions relative to signal yields. The $\gamma_{\Xi^+_c}$ values of $0.1$ and $0.5$ correspond to the blue and green bands, respectively.}
\label{fig:gamma-cp}
\end{figure}

\begin{figure}[H]
\begin{center}
\begin{overpic}[width=0.35\textwidth, trim=80 80 50 20,angle=0]{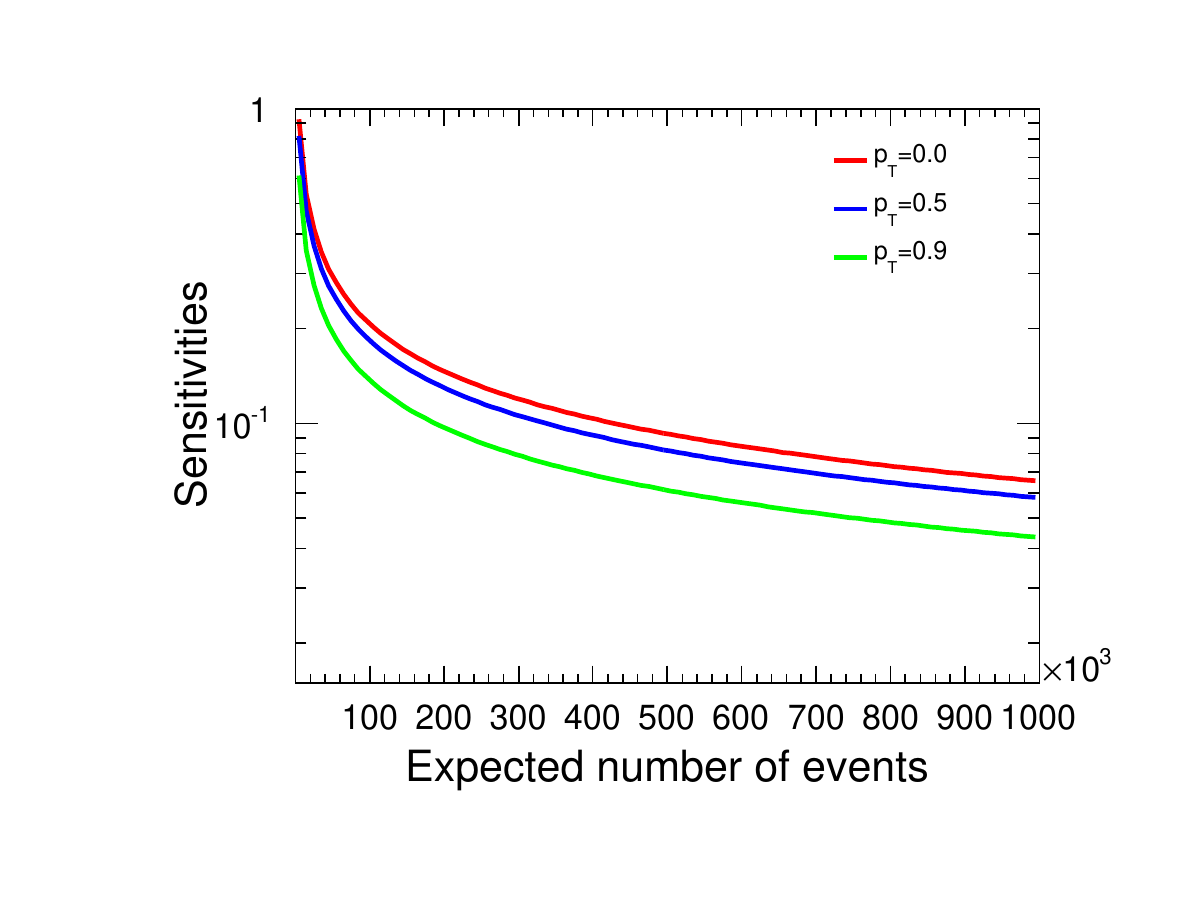}
\end{overpic}
\end{center}
\caption{The $\mathcal{C}_{CP}$ sensitivity distributions relative to signal yields assuming $\gammaXi=0.5$. The red, blue, and green lines represent the $p_{T}$ values of $0.0$, $0.5$, and $0.9$, respectively.}
\label{fig:alpha-cp-pt}
\end{figure}

Next we consider the sensitivity estimation on $CP$ violation parameter according to the error propagation equation \eq{delta-Acp}. The $CP$ violation in charm system should be more weak than that in bottom system~\cite{LHCb:2025ray}, so we constraint the upper limit of the parameters like  $\mathcal{A}_{CP}<0.05$ and the plot for the sensitivity of this parameter is shown in Fig.~\ref{fig:alpha-cp} with using the bands to display uncertainty. The statistical significances for $P$ and $CP$ violations are positively correlated, but the latter one requires more events to reach same sensitivities. For example, we find the parity sensitivity is $0.01$ when $N=300000$, while $CP$ sensitivity is about $0.04$ by comparing lines in Fig.~\ref{fig:delta-alpha} and Fig.~\ref{fig:alpha-cp} for $\alpha_{\Xi^+_c}=0.5$. The variations of  $\mathcal{B}_{CP}$ and $\mathcal{C}_{CP}$ sensitivities on different parameters are shown in Fig.~\ref{fig:beta-cp} and Fig.~\ref{fig:gamma-cp} respectively. In the last figure Fig.~\ref{fig:alpha-cp-pt}, we also show that the increasing of the polarization reduce the required events under the same significance for $\mathcal{C}_{CP}$ sensitivity.

\section{Summary}

The Cabibbo-favored three-body decay processes $\Xi_c^{+} \to \Xi^0 \pi^+\pi^0$ and $\Xi_c^{+} \to \Xi^- \pi^+\pi^+$ represent the most probable decay modes of $\Xi_c^{+}$. 
The intermediate processes $\Xi_c^+ \to \Xi(1620)^0 \pi^+\to\Xi^-\pi^+\pi^+$ and $\Xi_c^+ \to \Xi(1690)^0 \pi^+\to\Xi^-\pi^+\pi^+$ can also be measured with same method. By measuring the asymmetry decay parameters, the $P$ and $CP$ symmetries can be systematically tested. With sufficiently large data samples, the precision of the decay parameters $\alpha_{\Xi_c^{+}}$, $\beta_{\Xi_c^{+}}$, and $\gamma_{\Xi_c^{+}}$ can be significantly improved. Notably, for a given data sample, the measurement precision increases as the decay parameters approach 1.
Experimentally, there are some challenges in achieving controllable beam polarization. However, the effect of beam polarization can significantly enhance the precisions of $\gamma_{\Xi_c^{+}}$ and $\mathcal{A}_{CP}$ measurements. Therefore, it is very important to achieve a controllable beam polarization in future experiments, especially in the precise study of $CP$. For instance, when the phase difference $\Delta_{1}$ is close to $\pi/2$, and decay parameters are close to 1, the sensitivity becomes more obvious.
The expected $CP$ violation in $\Xi_c^{+}\rightarrow\Xi^-\pi^+\pi^+$ decay from weak interactions is on the order of $10^{-4}$ to $10^{-5}$, which is suppressed respecting to a rough benchmark number $10^{-3}$~\cite{Bianco:2003vb}, and mesurement for $\mathcal{A}_{CP}$ and $\mathcal{B}_{CP}$ will provide a good constraint to the total $CP$ violation in this decay. Therefore, it is estimated that at least $1.0\times10^9$ $\Xi_c^{+}\bar{\Xi}_c^{-}$ events are required to observe $CP$ violation, under the assumption of $\alpha_{\Xi^+_c} = 0.5$. The process could serve as an excellent probe to search for new sources of $CP$ violation beyond Standard Model. In this study, the sensitivities of the asymmetry parameters and $CP$ for the decay $\Xi_c^{+} \to \Xi^- \pi^+\pi^+$ are evaluated under various data sample sizes and beam polarization conditions. 
This work marks the first investigation of $P$ and $CP$ violation in the three-body decay of $\Xi_c^{+}$, providing essential theoretical support for future experiments, such as STCF.

\end{document}